\documentclass[%
reprint,
amsmath,amssymb,
aps,
]{revtex4-2}
\usepackage{xcolor}
\usepackage{graphicx}
\usepackage{dcolumn}
\graphicspath{{./Figures/}{./}}
\usepackage{enumerate}
\usepackage{subcaption}
\usepackage{bm}

\begin{document}
	
\preprint{APS/123-QED}
	
\title{Primary hemostasis and dynamics of clot formation after microvascular injury} 
	
\author{Alper Topuz, Gerhard Gompper, and Dmitry A. Fedosov}
\affiliation{Theoretical Physics of Living Matter, Institute for Advanced Simulation, Forschungszentrum J\"ulich, 52425 J\"ulich, Germany \\ 
Email: d.fedosov@fz-juelich.de} 
	
\date{\today}
	
\begin{abstract}
Primary hemostasis is initiated by platelet adhesion and aggregation at a site of vascular injury and is strongly regulated by 
local hydrodynamic conditions. At elevated shear rates, platelet capture is mediated by von Willebrand factor (vWF), a multimeric 
protein that undergoes shear-induced unfolding and becomes adhesive. We investigate early-stage clot formation under physiological 
high-shear-flow conditions by employing particle-based mesoscale hydrodynamics simulations with explicitly resolved red blood cells, 
platelets, and mechano-sensitive vWF in a microchannel geometry. The model incorporates vWF-mediated adhesion of platelets to a hemostatic 
surface, together with non-periodic inflow–outflow boundary conditions that allow continuous material supply and transport. We analyze 
the dynamics of platelet–vWF aggregation, clot growth dynamics, clot geometry and internal stresses, and thrombo-embolization across 
a range of elevated flow rates. Our results demonstrate that clot formation proceeds through the establishment of platelet–vWF aggregates 
at the hemostatic site, and that the clot reaches a finite size determined solely by hydrodynamic forces, without invoking biochemical 
stabilization mechanisms. Beyond a critical size, increased drag from fluid flow leads to recurrent embolization events that limit 
further growth. These findings highlight the central role of hydrodynamic stresses in regulating primary hemostasis and provide 
a mechanistic framework for understanding clot stability under physiological flow conditions.
\end{abstract}
	

\maketitle


\section{Introduction}
\label{sec:intro}

Hemostasis, or blood clotting, is the physiological response to vascular injury that prevents excessive blood loss through the formation 
of a localized clot. Its initial phase, referred to as primary hemostasis, is governed by platelet adhesion and aggregation at the injury 
site \cite{Monroe_MPC_2006,Versteeg_NFH_2013}. At low shear rates, platelet attachment is mainly mediated by integrin receptors. However, 
at shear rates exceeding approximately $600$–$900\ \mathrm{s}^{-1}$ \cite{Savage_IPA_1996,Jackson_PTF_2009}, which are typical for 
microvascular flows \cite{Popel_MH_2005,Lipowsky_MVR_2005}, integrin-mediated adhesion becomes insufficient. Under these conditions, 
platelet capture is facilitated by the multimeric protein von Willebrand factor (vWF) 
\cite{Savage_SSI_1998,Mendolicchio_NPW_2005,Reininger_MPA_2006,Ku_Card_2012}.

vWF is composed of dimeric subunits and can reach contour lengths of several tens of micrometers. It circulates freely in plasma and 
is also stored in platelets and endothelial cells \cite{Fowler_VWF_1985,Ceunynck_VWF_2013,Springer_VWF_2014}. Under quiescent or weak-flow 
conditions, vWF adopts a compact globular conformation that hides its platelet-binding domains \cite{Ulrichts_A1_2006,Springer_BPF_2011}. 
Exposure to elevated shear rates ($\gtrsim 2000\ \mathrm{s}^{-1}$) induces vWF stretching, as demonstrated in experiments 
\cite{Schneider_SIU_2007,Fu_VWF_2017} and simulations \cite{Katz_FIU_2006,Katz_DIP_2008,Hoore_FIA_2018}. This mechanical unfolding 
exposes adhesive domains that enable platelet binding. Single-molecule studies have further shown that stretched vWF can capture platelets 
when tethered to a surface \cite{Fu_VWF_2017}. At high shear rates, reversible platelet–vWF aggregates form both on surfaces and 
in suspension \cite{Schneider_SIU_2007,Chen_BCI_2013,Huck_VWF_2014,Schneider_BMC_2020}, and constitute the fundamental building blocks 
of primary hemostasis.

Platelets are anucleate cell fragments released from megakaryocytes, and play a central role in hemostasis \cite{Michelson_P_2007}. 
Upon vascular injury, platelets adhere to exposed ligands, such as collagen and/or vWF, undergo activation and shape change, and recruit 
additional platelets through chemical signaling \cite{Bryckaert_CML_2015,Savage_FSA_2002}. The resulting platelet plug forms within 
seconds, whereas subsequent activation and stabilization processes occur over longer timescales of minutes \cite{Hoekstra_Int_2019,Ku_BA_2022}. 
These disparate timescales have motivated a range of modeling approaches that focus on different phases of clot formation 
\cite{Nechipurenko_H_2020,Mukherjee_JTH_2024}.

Most existing computational studies focus on clot stabilization \cite{Mukherjee_SciRep_2024,Nechipurenko_BJ_2024,Hoekstra_PLOS_2023,Mukherjee_BMM_2021} 
or material transport within an already formed clot \cite{Mukherjee_JB_2021,Hoekstra_PLOS_2025,Karniadakis_PLOS_2020}. These approaches typically 
assume a prescribed clot geometry and neglect the presence of red blood cells (RBCs). Experimental studies, on the other hand, often target 
pathological occlusive thrombosis at extreme shear rates exceeding $10^4\ \mathrm{s}^{-1}$ \cite{Manning_BMES_2024,Ku_SIPA1_2022,Ku_BA_2022,Antaki_BJ_2022},
where shear-induced platelet aggregation dominates. In contrast, under physiological but elevated shear rates, the presence of soluble vWF 
is essential for clot growth, with a complex coupled dynamics of vWF, platelets and RBCs \cite{Schneider_BMC_2020}.

Dysregulation of primary hemostasis can lead to severe bleeding or thrombotic disorders. In high-shear environments, deficiencies in vWF 
cause von Willebrand disease \cite{Schneider_TH_2014}, while defects in the platelet GPIb receptor complex lead to Bernard–Soulier 
syndrome \cite{Lopez_BS_1998}. More recently, thrombo-inflammatory responses associated with COVID-19 have been linked to excessive 
exposure of anchored vWF, resulting in hypercoagulation and thrombo-embolism \cite{Choudhary_COVID_2021}. A quantitative understanding 
of early-stage clot formation under flow is therefore of both physiological and clinical importance.

In this work, we investigate primary hemostasis following vascular injury using particle-based hydrodynamics simulations with explicitly 
resolved RBCs, platelets, and shear-activated vWF in a microchannel under physiological high-shear conditions. The model incorporates 
mechano-sensitive vWF adhesion to both platelets and a hemostatic surface (or wound), together with non-periodic in- and out-flow boundary 
conditions that facilitate continuous material supply and transport to a growing clot. We analyze clot growth, clot geometry and internal 
stresses, structural stability, and thrombo-embolization, and demonstrate that hydrodynamic forces alone impose an upper bound on clot size.
Throughout this study, we consider three different flow rates, which result in wall-shear rates within the range between $1000\ \mathrm{s}^{-1}$ 
and $2500\ \mathrm{s}^{-1}$, representing physiological conditions in the elevated-shear-rate regime. Our results show that the dynamics of 
clot growth strongly depends on flow rate within a vessel, since it controls the rate of delivery of new hemostatic material. Interestingly, 
the steady-state size of the clot is nearly independent of the flow rate in the absence of platelet activation (i.e., no dynamic changes in the 
strength of involved bonds). The final size of the clot is governed by embolization events (i.e., shedding of clot fragments at the trailing edge) 
through the balance between the embolization and the deposition of new hemostatic material. Our simulations highlight the essential role 
of hydrodynamic interactions and shear-dependent mechanisms in regulating clot growth, stability, and embolization, providing a mechanical 
description of primary hemostasis.

\begin{figure*}
  \centering
  \includegraphics[width = \linewidth]{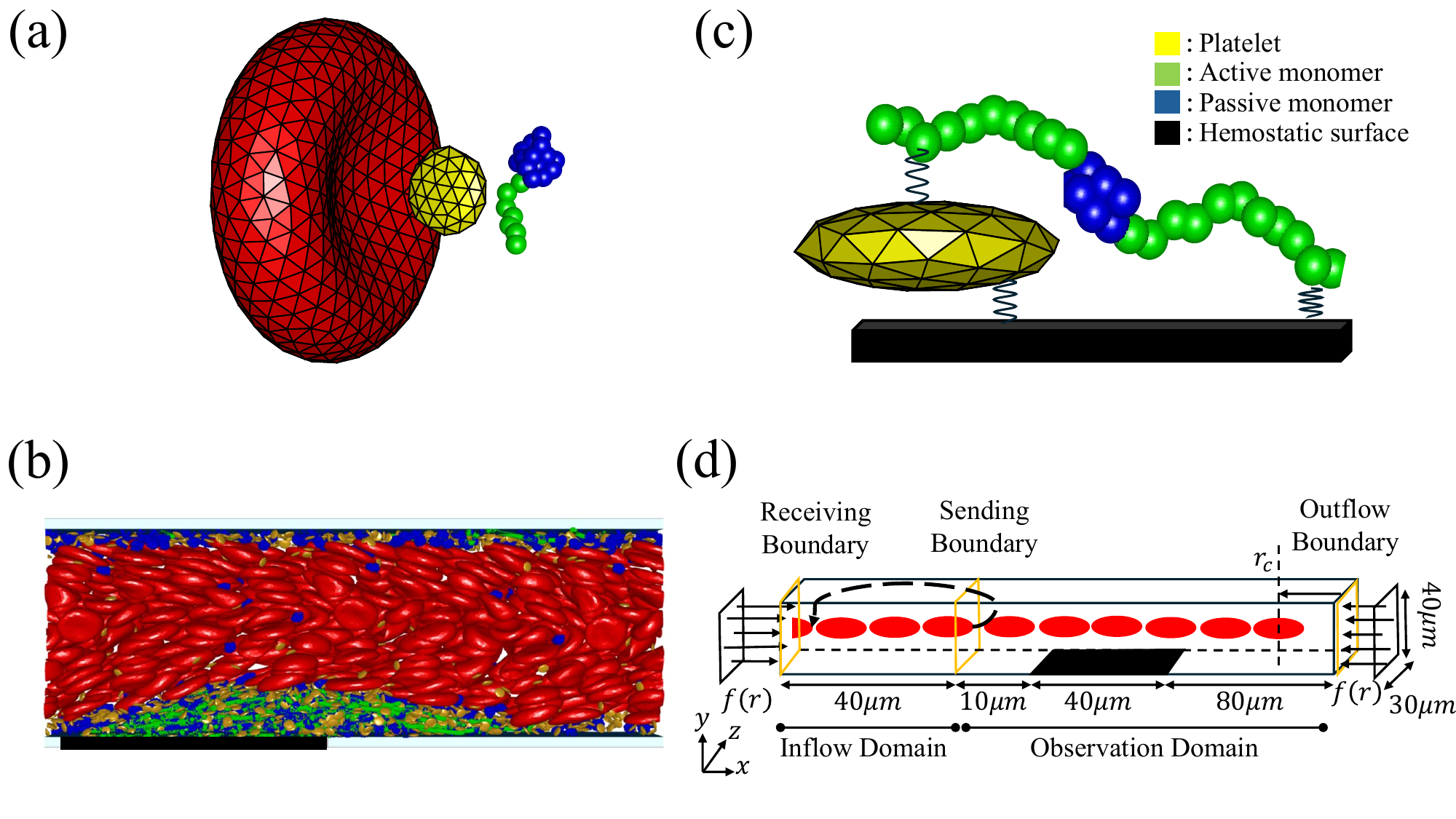}\\
  \caption{Illustration of the employed models and simulation setup. (a) Membrane models of the RBC (red) and platelet (yellow)
  together with a shear-activated vWF model (green for activated monomers and blue for inactivated monomers). (b) A snapshot 
  illustrating the growing clot. Same color code as in panel (a) applies here. The black bar represents a hemostatic surface, where 
  adhesive ligands are placed. (c) An illustration of various bonds between different hemostatic 
  components. (d) A schematic of the simulation domain of size 
  $170\mu m\times40\mu m\times30\mu m$. The domain is divided into an observation domain and periodic inflow domain. The inflow 
  domain assumes periodicity between its both ends, and constantly feeds the observation domain with flowing blood. The observation 
  domain is subject to outflow boundary conditions at its right end.}
  \label{fig:model}
\end{figure*}

\section{Models and methods}

Simulations of primary hemostasis require models of blood agonists (i.e., RBCs, platelets), vWF molecules, 
non-periodic boundary conditions and adhesive bonds. Figure~\ref{fig:model} illustrates these models and shows a simulation snapshot 
together with a channel setup whose detailed description is provided below. Fluid flow is simulated using the angular-momentum conserving 
smoothed dissipative particle dynamics (SDPD) approach, a mesoscopic method that discretizes the Navier–Stokes equations in a Lagrangian 
framework, while also incorporating thermal fluctuations in a thermodynamically consistent manner \cite{Espanol_SDPD_2003, Mueller_SDPD_2015, Quesada_CSF_2009}. 
Further details about fluid flow modeling and boundary conditions can be found in Appendix \ref{sec:app_model}.

\begin{table*}
\centering
\begin{tabular}{|c|c|c|c|c|} 
\hline
   & \multicolumn{2}{c|}{RBC} & \multicolumn{2}{c|}{Platelet} \\
 \hline
  Parameters &  Scaled units & Physical units &  Scaled units & Physical units \\
\hline
$N_v$      & $500$  & &  $60$ &    \\
$A$      &     & $132.83 \times 10^{-12}$ $\text{m}^2$ &  $0.162 D_r^2$ & $6.85 \times 10^{-12}$ $\text{m}^2$ \\
$V$      & $0.336$ $D_r^3$      & $92.4 \times 10^{-18}$ $\text{m}^3$  & $0.0041$ $D_r^3$ & $1.127 \times 10^{-18}$ $\text{m}^3$  \\
$\mu$    & $4.77\times 10^4$ $k_BT / D_r^2$  & $4.83 \times 10^{-6}$ $\text{N/m}$ & $4.77\times 10^5$ $k_BT / D_r^2$  & $4.83 \times 10^{-5}$ $\text{N/m}$ \\
$\kappa$ & $70$ $k_BT$ & $3 \times 10^{-19}$ $\text{J}$ & $700$ $k_BT$ & $3 \times 10^{-18}$ $\text{J}$ \\
$k_d$    & $4.23\times 10^4$ $k_BT / D_r^2$  & $4.28 \times 10^{-6}$ $\text{N/m}$ & $4.23\times 10^5$ $k_BT / D_r^2$  & $4.28 \times 10^{-5}$ $\text{N/m}$ \\
$k_a$    & $2.07\times 10^6$ $k_BT / D_r^2$  & $2.1 \times 10^{-4}$ $\text{N/m}$  & $3.81\times 10^6$ $k_BT / D_r^2$  & $3.86 \times 10^{-4}$ $\text{N/m}$ \\
$k_v$    & $1.37\times 10^7$ $k_BT / D_r^3$  & $213.28$ $\text{N/m}^2$ & $2.75\times 10^7$ $k_BT / D_r^3$  & $428.11$ $\text{N/m}^2$ \\
\hline
\end{tabular}
\caption{Parameters for RBCs and platelets are reported in units of the effective RBC diameter $D_r = \sqrt{A_r/\pi}$ and the thermal 
energy $k_B T$ at $T = 310,\text{K}$, together with their corresponding physical values. Here, $N_v$ denotes the number of membrane vertices, 
$A$ the membrane surface area, $V$ the enclosed volume, $\mu$ the membrane shear modulus, and $\kappa$ the bending rigidity. The parameters 
$k_d$, $k_a$, and $k_v$ represent the coefficients enforcing local area, global area, and volume constraints, respectively. Throughout all 
simulations, we set $A_r = 132.83$ and $k_B T = 0.1$, so that the effective diameter becomes $D_r = 6.5$.}
\label{tab:rbc}
\end{table*}

\subsection{RBCs and platelets}

Both RBCs and platelets are represented as deformable particles, with a membrane enclosing a fluid volume. The membrane is modelled 
as a triangulated network consisting of $N_v$ vertices with positions $\mathbf{x}_{i}$ where $i = 1\dots N_v$, as illustrated in 
Fig.~\ref{fig:model}(a) \cite{Fedosov_RBC_2010, Fedosov_SCG_2010, Noguchi_STV_2005}. These vertices are interconnected by elastic springs 
that define the cell surface geometry. The total elastic energy of the network is
\begin{equation}
U(\{{\bf x}_i\}) = U_s + U_b + U_a + U_v,   
\end{equation}
where $U_s$ accounts for in-plane stretching due to spring deformation, and $U_b$ represents the bending resistance of the membrane. 
The terms $U_a$ and $U_v$ enforce surface area and volume conservation, respectively. These constraints reflect key mechanical properties 
of biological membranes: area incompressibility of the lipid bilayer and incompressibility of the enclosed cytoplasmic fluid. 
A more detailed formulation of these energy terms for RBCs can be found in Refs.~\cite{Fedosov_SCG_2010, Fedosov_RBC_2010}.

Mechanical behavior of the model membranes is characterized by two primary elastic parameters: the shear modulus $\mu$, which 
governs in-plane shear elasticity, and the bending rigidity $\kappa$, which determines resistance to bending. Parameters used for RBCs 
and platelets are summarized in Table~\ref{tab:rbc}, given in both simulation and physical units. Platelets are modeled as nearly 
rigid oblate spheroids with a diameter of 2 $\mu m$ and an aspect ratio of $0.3$. To represent their reduced deformability compared 
to RBCs, the membrane rigidity of platelets is set approximately ten times larger than that of RBCs. Furthermore, a local spontaneous 
curvature is imposed on the platelet surface to maintain their oblate geometry. In contrast, RBCs are modeled without spontaneous 
curvature.

\subsection{Shear-activated vWF}

vWF is a self-attractive protein that becomes functionally adhesive as it is stretched in response to mechanical forces. Under 
low shear stress or in the absence of flow, vWF adopts a compact, globular conformation that prevents it from binding to platelets. 
However, when exposed to sufficiently high shear stresses, it unfolds and exposes adhesive domains, enabling platelet capture 
\cite{Schneider_SIU_2007, Reininger_VWF_2008, Fu_VWF_2017}. Our vWF model is a bead-spring polymer chain consisting of $N_m = 30$ 
self-avoiding, attractively interacting monomers [see Fig.~\ref{fig:model}(a, c) and Appendix \ref{sec:app_vwf}], following previous 
approaches developed in Refs.\cite{Katz_FIU_2006, Katz_DIP_2008, Katz_SEU_2007}. The activation of vWF beads for adhesion is governed 
by the local conformation of vWF chain \cite{Huisman_ADAM_2017} (see Appendix \ref{sec:app_vwf} for details). Activated vWF monomers 
can adhere to platelets and ligands placed at the hemostatic surface, see Fig.~\ref{fig:model}(c).

\subsection{Adhesion interactions}

The major adhesive interactions in primary hemostasis at elevated shear stresses arise between vWFs, platelets, and the hemostatic surface. 
In this study, we model the adhesive bonds between these agonists without any activation of platelets. We model the various adhesive 
interactions using discrete bonds, which can dynamically form and break depending on various factors.  

The probability of bond rupture usually increases with increasing strength of the applied forces. This bond behavior is 
well described by a {\em slip-bond} model, which assumes a positive correlation between the rupture probability and the applied 
force. Consequently, a high rupture probability means a short lifetime of the bond, such that the lifetime of a slip bond decreases 
with increasing applied force. However, a number of experiments \cite{Dembo_RLK_1988,Kong_DCB_2009,Thomas_BAC_2002} have 
demonstrated that the lifetime of certain ligand-receptor interactions may increase as the applied force is elevated. 
In this case, the bond is referred to as a {\em catch bond} \cite{Dembo_RLK_1988}. Clearly, any physical bond will eventually 
rupture at large enough forces. Therefore, the catch behavior of a bond should rather be considered as a dual {\em catch-slip}
behavior, where the lifetime first increases (i.e., catch behavior) and then decreases (i.e., slip behavior) with 
increasing applied force. Even though the catch-slip behavior has been found only for a few ligand-receptor pairs 
\cite{Kong_DCB_2009,Thomas_BAC_2002}, this type of bonds is well accepted in the context of biological systems \cite{Thomas_BAC_2002}.
For example, adhesion of vWF to platelet GPIb$\alpha$ receptor exhibits the catch-slip behavior \cite{Doggett_VWF_2002,Yago_PGP_2008,Kim_MSB_2010}.

\begin{figure}
  \centering
  \includegraphics[width=1\linewidth]{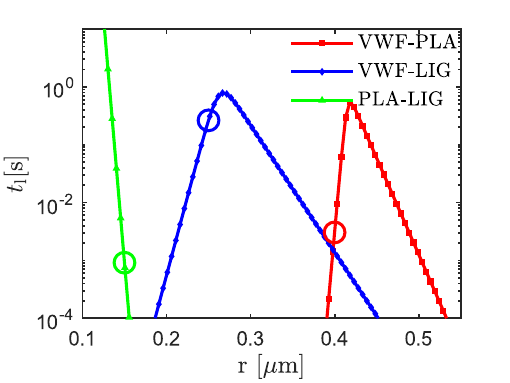}
  \captionsetup{format=plain}
  \caption{Comparison of lifetimes $t_l = 1/k_{\rm off}$ for different bond types of vWF-platelet (VWF-PLA), vWF-hemostatic 
  surface ligand (VWF-LIG), and platelet-hemostatic surface ligand (PLA-LIG) interactions (see Table \ref{tab:bond}), as a function 
  of bond length $r$. 
  The circles on each curve indicate the equilibrium bond length of the adhesive interaction represented by a harmonic spring.}
  \label{fig:lifetime}
\end{figure}

Attachment of platelets to vWF is modeled with the catch-slip mechanism described above. Another pathway for platelet recruitment from the blood 
is the direct binding of platelets to exposed ligands at the damaged endothelial surface. This interaction, however, mediates the platelet capture at 
moderate to low shear rates and its effect at high shear rates is small. Therefore, we model platelet interaction with a hemostatic surface via a 
slip-bond model, with a lower bond lifetime compared to vWF-platelet interactions. Activated vWFs can bind to a hemostatic surface through 
various surface ligands at the damaged endothelial surface. We describe the adhesive interaction between vWF and ligands at the hemostatic 
surface by a catch-slip bond model, but keep bond lifetime highest among the other interactions in order to maintain a strong adhesion 
of vWF to the surface, compare Fig.~\ref{fig:lifetime}. 

Adhesive interactions between activated vWF monomers, platelet membrane vertices and hemostatic surface are modeled via the formation 
of reversible bonds, characterized by predefined association and dissociation rates, $k_{\rm on}$ and $k_{\rm off}$, respectively. 
Once a bond is established, it is represented by a harmonic potential $U_b(r) = k_b (r - r_0)^2$, where $k_b$ is the bond stiffness 
of the corresponding adhesive interaction, and $r_0$ is the equilibrium bond length. The probabilities of bond formation ($P_{\rm on}$) 
and rupture ($P_{\rm off}$) are governed by equations, which relate these probabilities to the corresponding 
kinetic rates \cite{Bell_MSA_1978},
\begin{eqnarray}
\frac{{\rm d}P_{\rm on}}{{\rm d}t} &= - k_{\rm on} P_{\rm on}, \quad \text{for } r \leq r_{\rm cut}^{\rm on}, \\
\frac{{\rm d}P_{\rm off}}{{\rm d}t} &= - k_{\rm off} P_{\rm off}, \quad \text{for } r \leq r_{\rm cut}^{\rm off},
\end{eqnarray}
where $r_{\rm cut}^{\rm on}$ and $r_{\rm cut}^{\rm off}$ are the cutoff ranges for bond association and dissociation, respectively. 
Note that $P_{\rm on} = 0$ for $r > r_{\rm cut}^{\rm on}$ and $P_{\rm off} = 1$ for $r > r_{\rm cut}^{\rm off}$. In simulations, 
we assume $k_{\rm on}$ to be constant, while $k_{\rm off}$ follows the aforementioned catch-slip or slip behavior. This is described 
by the two-pathway model \cite{Pereverzev_TPM_2005}, which assumes two force-dependent barriers for bond dissociation,
\begin{equation}
k_{\rm off} = k^{\rm 0}_{\rm c}\exp{\left( \frac{\lambda_c (r-x_{\rm eq}) \delta_{\rm c} }{k_{\rm B}T} \right)} + 
k^{\rm 0}_{\rm s}\exp{\left(\frac{\lambda_s (r-x_{\rm eq}) \delta_{\rm s}}{k_{\rm B}T}\right)} \mbox{,}
\label{EQtwopath}
\end{equation}
where the first term represents a catch-bond dissociation rate, while the second term corresponds to a slip-bond rate. 
Here, $k^{\rm 0}_{\rm c}$ and $k^{\rm 0}_{\rm s}$ are the catch and slip equilibrium off-rates. $\lambda_c$ and $\lambda_s$ 
represent the strengths of the catch and slip parts, and $\delta_{\rm c} = x_{\rm c}-x_{\rm eq}$ and $\delta_{\rm s} = x_{\rm s}-x_{\rm eq}$, 
where $x_{\rm c}$, $x_{\rm s}$, and $x_{\rm eq}$ are the catch, slip, and equilibrium characteristic lengths. Note that the condition 
$x_{\rm c} < x_{\rm eq} < x_{\rm s}$ should be satisfied, so that the catch part dominates for $r < x_{\rm eq}$, while 
the slip part dominates for $r>x_{\rm eq}$. 

\begin{table*}
\centering
\begin{tabular}{|c|c|c|c|c|c|c|c|c|c|} 
 \hline
 bond type &  $k^{\rm 0}_{\rm c}$ & $x_c$ & $x_{\rm eq}$ & $k^{\rm 0}_{\rm s}$  & $x_s$
 & $r_{\rm cut}^{on}$ & $r_{\rm cut}^{\rm off}$ & $k_{\rm b}$ & $r_{0}$ \\
\hline
$\mathrm{vWF - PLA}$    & $5.75 \times 10^{-1} / \tau$  & $0.0385D_r$ & $0.0615 D_r$ & $5.75 \times 10^{-4} / \tau$ &  $0.0662D_r$ &$0.067D_{r}$&$0.23D_{r}$ &$8.45 \times 10^{6} k_{\rm B}T / D_{r}^2$ &$0.0615D_{r}$\\
$\mathrm{vWF - LIG}$    & $5.75 \times 10^{-3} / \tau$  & $0.0307D_r$ & $0.0384 D_r$ & $5.75 \times 10^{-4} / \tau$ &  $0.0415D_r$ &$0.046D_{r}$&$0.23D_{r}$ &$8.45 \times 10^{6} k_{\rm B}T / D_{r}^2$ &$0.0384D_{r}$\\
$\mathrm{PLA - LIG}$    & $0$  & $0$ & $0.023  D_r$ & $1.44 / \tau$ &  $0.037D_r$  &$0.0308D_{r}$&$0.23D_{r}$ &$2.11 \times 10^{6} k_{\rm B} T / D_{r}^2$ &$0.023D_{r}$\\
\hline
\end{tabular}
\caption{Parameters of the bond dissociation rate $k_{\rm off}$, see Eq.~\eqref{EQtwopath}, for the different bond types. In all cases, $\lambda_c = \lambda_s = 1.06 \times 10^5\,k_{\rm B}T / D_r^2$ and $k_{\rm on}=5.73\times 10^{3}/\tau$.}
\label{tab:bond}
\end{table*}

\subsection{Simulation setup} 

Blood flow is simulated within a channel of size $L_x \times L_y \times L_z = 170\mu m \times 40\mu m \times 30\mu m$, which consists 
of inflow and observation parts (see Appendix \ref{sec:app_model} for more details). The periodic inflow domain has a size of $L_x \times L_y \times L_z = 40\mu m \times 40\mu m 
\times 30\mu m$ and contains $180$ RBCs, $384$ platelets, and $192$ vWFs, which correspond to volume fractions of approximately $42 \%$ for
RBCs (hematocrit), $1.7 \%$ for platelets, and $0.8 for \%$ vWFs. 
Note that the simulated volume fractions 
of vWFs and platelets are higher than under physiological conditions, in order to reduce the time scale of hemostatic plug formation 
to a reasonable computation time. The inflow domain strictly conserves its number of particles due to periodicity, but the number of 
blood agonists can vary in the observation domain due to arrest of hemostatic material at the wound. The simulation domain is bounded 
by no-slip walls in the $y$ direction, has periodic boundary conditions in the $z$ direction, and non-periodic boundary conditions 
in the $x$ direction [see Fig.~\ref{fig:model}(c) and Appendix \ref{sec:app_model}]. The suspending fluid represented by SDPD particles has the number density of 
$n=16.875/ r_c^3$ with the cutoff radius $r_c = 0.23 D_r$ for particle interactions. Fluid viscosity is set to $\eta = 100$, which defines a time scale 
$\tau = \eta D_r/ \mu_r$. $\tau \approx 0.0016$ s for the blood plasma viscosity of $\eta = 1.2 \times 10^{-3}$ $Pa \cdot s$. 
The flow is driven by a pressure gradient of $\Delta P / L_x = fn = c k_BT / r_c^4$, where $f$ is the force applied to each SDPD 
fluid particle and $c$ takes the values of $81$, $121.5$, and $162$ for low-, mid- and high-flow rate simulations, respectively. 

Excluded-volume interactions between RBCs, platelets, and vWFs are implemented through the repulsive part of the LJ potential 
in Eq.~(\ref{eq:lj}). For all repulsive interaction pairs, $\epsilon=16k_{\rm B}T$ and $r_{LJ} = 2^{1/6} \sigma$. However,
the effective particle diameters are somewhat different due to slight differences in the discretization resolution of various 
suspended components, such that $\sigma =0.077 D_r$ for vWF-vWF interactions, $\sigma =0.062 D_r$ for RBC-vWF and platelet-vWF 
repulsion, and $\sigma =0.046 D_r$ for RBC-RBC, platelet-platelet, and RBC-platelet interactions.

In the simulations, platelet vertices, activated vWF monomers, and ligand particles placed at the hemostatic surface (or wound) 
can form adhesive bonds [see Fig.~\ref{fig:model}(d)]. Adhesive bonds are permitted only between platelet–vWF, platelet–wound, and 
vWF–wound pairs. We model the hemostatic surface at the lower wall by a cloud of stationary ligand particles, which forms a two-dimensional 
square lattice characterized by the lattice spacings $\Delta x = \Delta z = 0.25\,\mu\mathrm{m}$, corresponding to a surface number density 
of $16.23~\mathrm{particles}/\mu\mathrm{m}^2$. The overall hemostatic surface patch at the lower wall has dimensions 
$L_x \times L_z = 40\,\mu\mathrm{m} \times 30\,\mu\mathrm{m}$ [see Fig.~\ref{fig:model}(c)].

The parameters for the various bond types are given in Table \ref{tab:bond}. The corresponding lifetimes $t_l = 1/k_{\rm off}$ as 
a function of the bond length $r$ are displayed in Fig.~\ref{fig:lifetime}. The value for the on-rate $k_{\rm on} = 5.73 \times 10^{3}/\tau$ 
is constant and same for all bond types. The given set of on- and off-rates can produce reversible aggregation of the hemostatic plug, 
i.e., the clot is able to dissolve upon cessation of fluid flow. 

\begin{figure*}
  \centering
  \includegraphics[width=\textwidth]{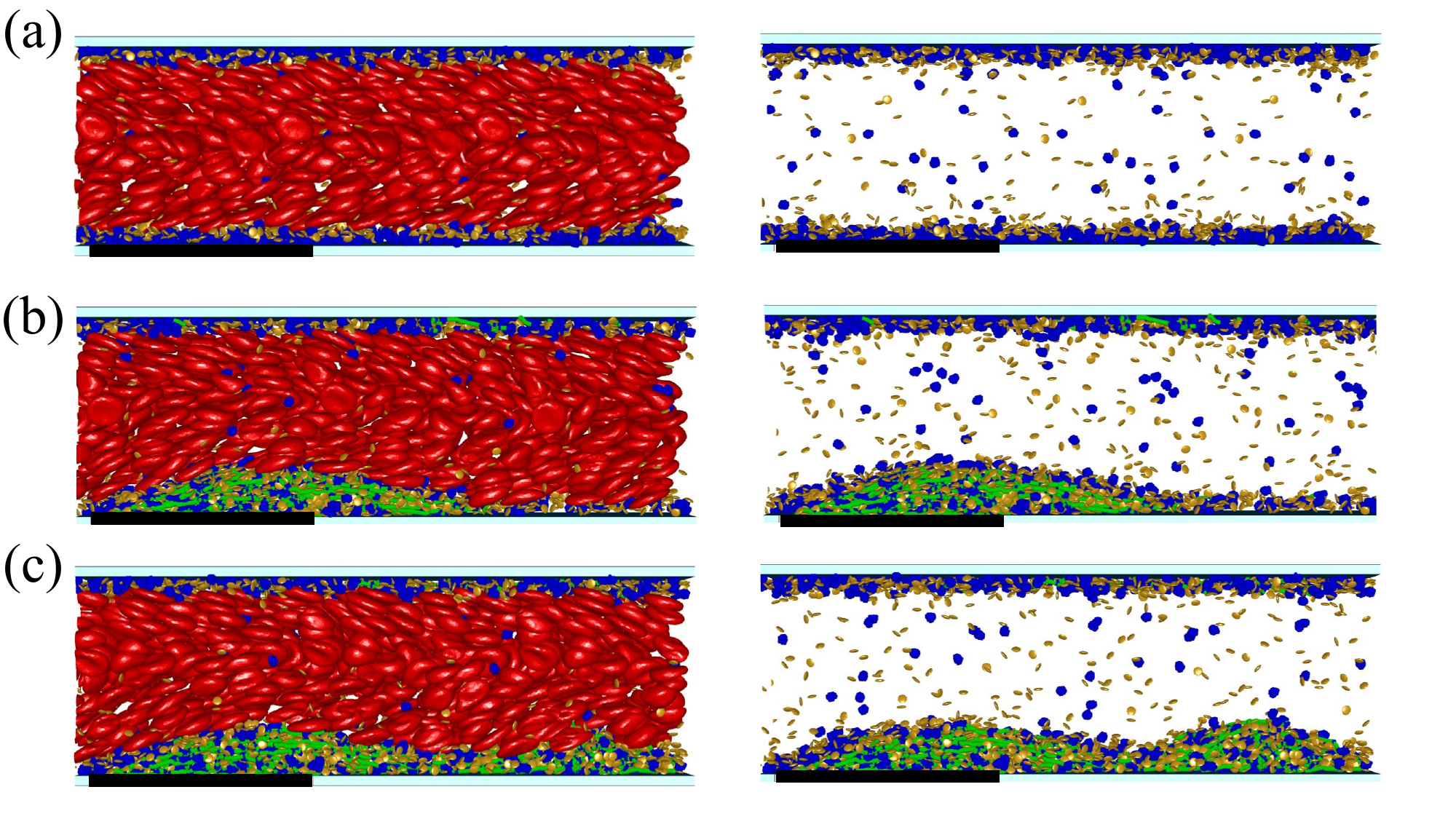}
  \captionsetup{format=plain}
  \caption{Simulation snapshots illustrating clot formation and development. The black bar depicts the location of the hemostatic surface in all pictures. (a) After marginating into RBC-free layer, the concentration of 
  platelets and vWFs near the wall is significantly increased. This snapshot 
  shows the initial condition obtained from a low-shear simulation. (b) As time progresses, adhered hemostatic material gradually 
  accumulates at the hemostatic surface. (c) Excess hemostatic material is pushed away from the wound region due to hydrodynamic stresses.}
  \label{fig:snapshots}
\end{figure*}

\section{Results}

\subsection{vWF-platelet aggregation and clot formation}

The formation of a vWF-platelet plug at a vascular injury requires high enough flow stresses that are able to stretch and activate vWF. 
We simulate three different scenarios to analyze the effect of flow rate on clot formation: the low-,  mid-, and high-flow rate 
simulations with the wall shear rates $\dot{\gamma}_w = fn L_y /(2 \eta) \approx 1152\ {\rm s}^{-1}$,  $\dot{\gamma}_w =1728\ {\rm s}^{-1}$, 
and $\dot{\gamma}_w = 2304 {\rm s}^{-1}$, respectively. These rates are relevant for the circulation in arterioles and small 
arteries \cite{Sakariassen_Thrombus_2015}). In principle, the low- and mid-flow rate scenarios have shear rates below the critical shear rate 
$\dot{\gamma}_c \approx 2000\ {\rm s}^{-1}$ for soluble vWF to extend \cite{Schneider_SIU_2007,Reininger_VWF_2008,Fu_VWF_2017}. However, 
quasi-confinement effects induced by flowing RBCs on vWFs near the wall reduce this critical shear threshold \cite{Rack_VWF_2017}. 
In all our simulations, vWF stretches and exposes its adhesive domains, initiating clot formation.

Figure~\ref{fig:snapshots} shows representative simulation snapshots for various stages of the clot-formation process. The left column 
includes all blood components, such as RBCs and hemostatic material, while the right column displays only the hemostatic material. At the start 
of the simulation, when the driving force is just switched on to initiate blood flow, all vWFs have a globular form (i.e., they are non-adhesive), 
and are homogeneously distributed within the simulation domain together with platelets and RBCs, without any bonded structures. Under flow, RBCs 
migrate toward the channel center, while platelets and vWF globules slowly migrate toward the channel walls. The migration into the vicinity of 
the walls is called margination, and has also been observed for leukocytes and drug delivery carriers in blood flow \cite{Goldsmith_MLB_1984,Fedosov_WBC_2012,Mueller_MPB_2014,Rack_VWF_2017}. The margination of platelets and vWFs is facilitated by the hydrodynamic lift force on deformable 
RBCs, which thereby effectively push the other blood components into a RBC-free layer (RBC-FL) near the wall. As a result, the concentration of platelets 
and vWFs in the RBC-FL strongly increases, providing favorable conditions for the formation of a blood clot. In a preliminary simulation, 
we do not allow formation of any bonded structures as the RBC-FL develops. Figure~\ref{fig:snapshots}(a) shows a simulation snapshot with 
marginated platelets and vWFs within the RBC-FL, which serves as an initial condition for hemostatic simulations. 

A typical hemostatic reaction is initiated through the adhesion of vWFs at different locations of the injury site, with a subsequent capture 
of flowing platelets. The adhered vWFs form small "islands" of attached material, which grow into three-dimensional structures, and eventually 
merge into a single clot over different timescales, depending on the flow rate. Figure~\ref{fig:snapshots}(b) shows a snapshot when these small 
islands have merged into one intact, connected blood clot. Once the clot reaches a certain size, it occludes certain part of the channel and 
experiences increased drag forces due to hydrodynamic stresses. These stresses impose an upper limit on the clot size, since further growth 
causes a structural instability, leading to thrombo-embolization. Figure~\ref{fig:snapshots}(c) displays a typical conformation at the onset of 
embolization. Shortly after this snapshot, a small aggregate detaches from the main clot, and enters the circulation.

Further, we analyze clot formation and dynamics for different flow rates. Specifically, we focus on geometric properties, dynamics of the clot, 
bond formation characteristics, and various time scales occurring in the hemostatic process. 

\begin{figure*}
  \centering
  \includegraphics[width=\textwidth]{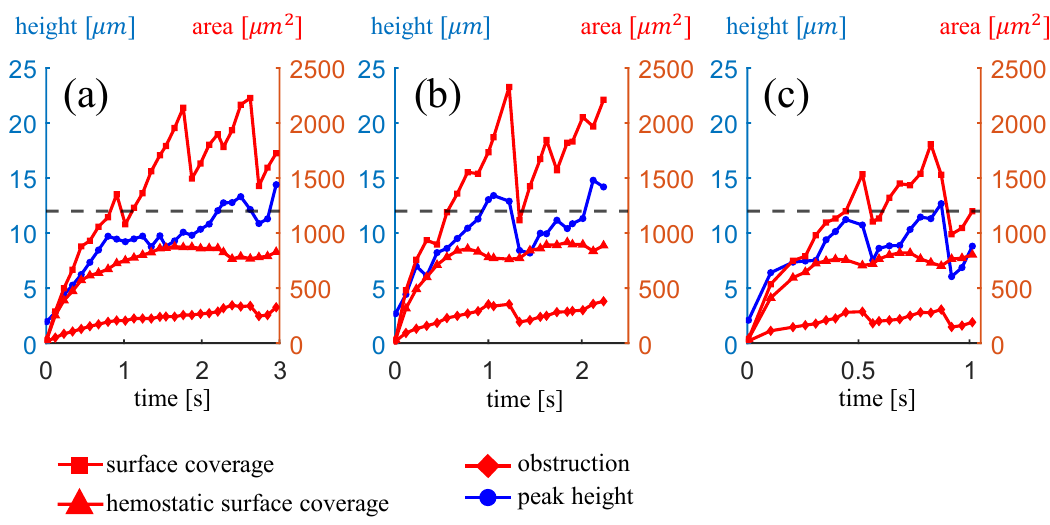}
  \captionsetup{format=plain}
  \caption{Time-dependent total surface coverage, hemostatic surface coverage, obstruction area, and peak height of the vWF-platelet aggregate 
  for (a) low-flow rate, (b) mid-flow rate, and (c) high-flow rate. The corresponding coverage areas are obtained by projecting the individual 
  platelets onto the lower wall (total surface coverage), onto the wound area (hemostatic surface coverage), and onto the plane perpendicular 
  to the flow direction (obstruction area). The peak height corresponds to the maximum instantaneous y-coordinate of the platelets. The dashed 
  line displays the total area of the wound (or hemostatic surface), corresponding to $1200 \mu m^2$.}
  \label{fig:surfaceCoverage}
\end{figure*}

\subsection{Clot geometry -- surface coverage, clot thickness, and embolization events}

The purpose of the hemostatic process is the cessation of bleeding upon vascular injury. This process involves several consecutive steps, the
first of which is the formation of a vWF-platelet plug at the wound. The relevant measure of this very first step is the surface 
coverage of the clot due to the attachment of platelets to the wound as a function of time. 

Figure~\ref{fig:surfaceCoverage} shows the different surface-coverage measures of the vWF-platelet plug for low-, mid- and high-flow-rate 
simulations. Here, the total surface coverage is computed by projecting the platelets onto the lower wall, hemostatic surface coverage
is calculated by computing the overlap of the total surface coverage with the wound area, and the obstruction area is calculated by 
projecting the platelets onto the plane perpendicular to the flow direction. In all cases, a typical time scale $\tau_{\rm e}$ for the 
hemostatic surface coverage can be identified by 
considering the exponential approach of saturation, i.e., $a\left(1 - \exp{(-t/\tau_{\rm e})}\right)$. 
The corresponding time required to attain $95 \%$ of the steady-state value is chosen as the steady-state time (i.e., $\tau_{\rm ss}$). 
These two measures of the typical time scale display essentially the same dependence on flow rate and other parameters. The steady-state times 
for the low-, medium- and high-flow-rate cases are $\tau_{\rm ss}=1.18 {\rm s}$, $\tau_{ss}=0.79 \rm s$, and $\tau_{ss}=0.41 \rm s$,
where the hemostatic surface coverages reach $833.4 \mu m^2$, $859.6 \mu m^2$, and  $774.1 \mu m^2$, respectively. Note that the total surface 
area of the wound is $1200 \mu m^2$ (dashed line in Fig.~\ref{fig:surfaceCoverage}) in all simulations, such that the formed clot is 
not able to cover the whole surface area of the wound. The steady-state times $\tau_{ss}$ exhibit an inversely proportional relation 
with the wall-shear rate (i.e, $\tau_{ss} \approx 1.3\times10^{3}/\dot{\gamma}_{w}$). The steady-state hemostatic surface coverage 
values reach approximately the same value in all simulations. 

Unlike hemostatic surface coverage, the total surface coverage (i.e., the total area of a projected clot) is frequently interrupted by 
embolization (i.e., detachment) events. It was not possible to reliably compute the time between two consecutive embolizations, since 
only $2-3$ such events occur during the typical simulation time. When an embolization event happens, the fragment detaching from the clot does not 
overlap with the wound area. This can clearly be seen from Fig.~\ref{fig:surfaceCoverage}, as the drops in the total surface coverage 
do not affect the hemostatic surface coverage. This implies that once the platelet plug is formed at the wound, the clotting process 
can proceed with secondary hemostasis without any interruption from the embolizations.  

Figure \ref{fig:surfaceCoverage} also shows the obstruction area (i.e., the area of the clot obtained by projecting the individual 
platelets onto the plane perpendicular to the flow direction) and the peak height (i.e., maximum $y$-coordinate of all platelets) 
of the clot. The drag force on the clot is proportional to the obstruction area or equivalently to the peak height. The hydrodynamic 
stresses on the clot increase as the channel becomes increasingly occluded. Beyond some critical value, the drag force becomes sufficient 
to scavenge excess clotting material from the hemostatic plug, thereby limiting further growth of the clot. Note that, for the low-flow-rate case, peak height remains nearly constant after the steady state has been reached, even though there are repeated detachments from 
the clot. This is relevant for the embolization mechanism that will be discussed below. For the mid- and high-flow-rate cases, the peak 
height value is subject to sudden drops as detachments take place. The maximum peak heights for the low-, mid- and high-flow-rate simulations 
are $14.72 \mu m$, $14.79 \mu m$, and $12.67 \mu m$, respectively. The corresponding obstruction areas are $388 \mu m^2$, $379 \mu m^2$, 
and $305 \mu m^2$, respectively, see Fig.~\ref{fig:surfaceCoverage}.

From the simulation results for the peak heights, obstruction areas, hemostatic and total surface coverage areas, we conclude that 
clots subjected to different flow rates reach approximately the same geometric measures, but at different timescales. Clots at the 
highest flow rate remains slightly smaller than the clots at lower flow rates, due to high drag forces and hydrodynamic stresses. 
Therefore, an important result of our simulations is the emergence of an upper limit for the size of the blood clot. Although our model 
includes only hydrodynamics, bond formation, and the mechanics of blood agonists, it consistently yields a clot that does not occlude 
the entire channel. This suggests that invoking additional mechanisms, such as bio-chemical reactions that govern, for instance platelet 
activation, is not necessary to achieve a finite clot size.

\begin{figure*}[t]
  \centering
  \includegraphics[width=\textwidth]{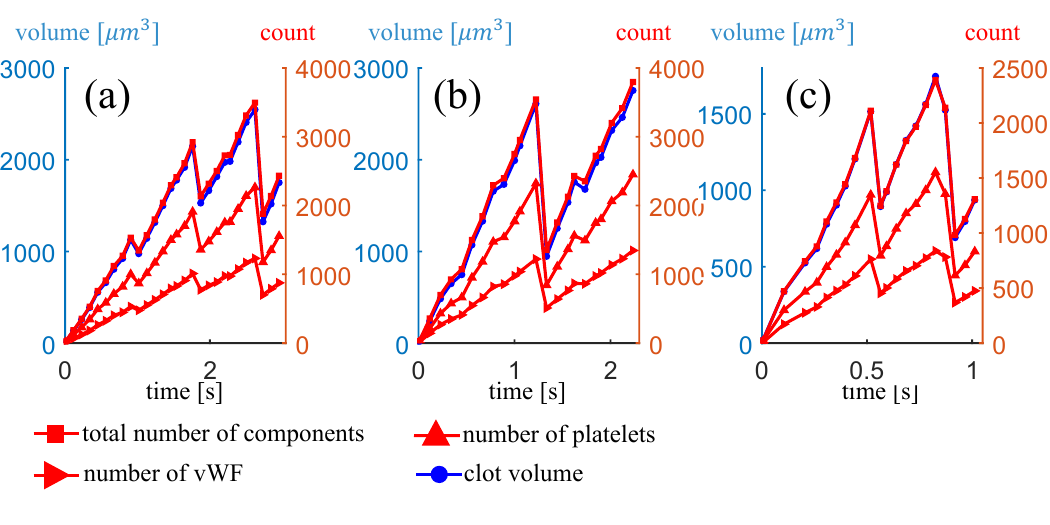}
  \captionsetup{format=plain}
  \caption{Time-dependent measure of a number of platelets, vWFs, total number of components, and total volume of the platelet-vWF 
  plug for (a) low-flow rate, (b) mid-flow rate, and (c) high-flow rate. The volume is calculated using the total sum of individual 
  platelet volumes. The total number of components corresponds to the total count of vWFs and platelets.}
  \label{fig:relativeCounts}
\end{figure*}

\subsection{Clot structure -- vWF and platelet content, and embolization events}

Figure \ref{fig:relativeCounts} shows the counts of platelets and vWFs in the clot, and the total volume of the platelets for different 
flow rates. The embolization events can easily be traced by the sudden drops in the time dependence of these quantities. For all simulations, 
the ratio of the vWF and platelet counts within the clot is almost constant, and equals to about $0.55$ (i.e., about two platelets per vWF). 
Even when an embolization event takes place, this ratio does not change. This implies that the clot is homogeneous in terms of platelet and vWF counts. 
The ratio matches well the number fraction in the periodic inflow domain (i.e., $192$ and $384$ for vWF and platelets, respectively). 
We also measure the rate of material deposition by tracking the time derivative of the total number of components (i.e., sum of platelets 
and vWFs) in the clot, which yields deposition rates of $1907 \, s^{-1}$, $3192 \, s^{-1}$, and $4293 \, s^{-1}$ for the three flow rates. 
These values are proportional to the wall shear rate with the scaling $\dot{N} \approx 1.79 \dot{\gamma}_{\rm wall}$. Note that even if 
there are jumps in the curves in Fig.~\ref{fig:relativeCounts}, after an embolization the clot continues to grow with approximately 
the same rate.   

\begin{figure*}[t]
  \centering
  \includegraphics[width=\textwidth]{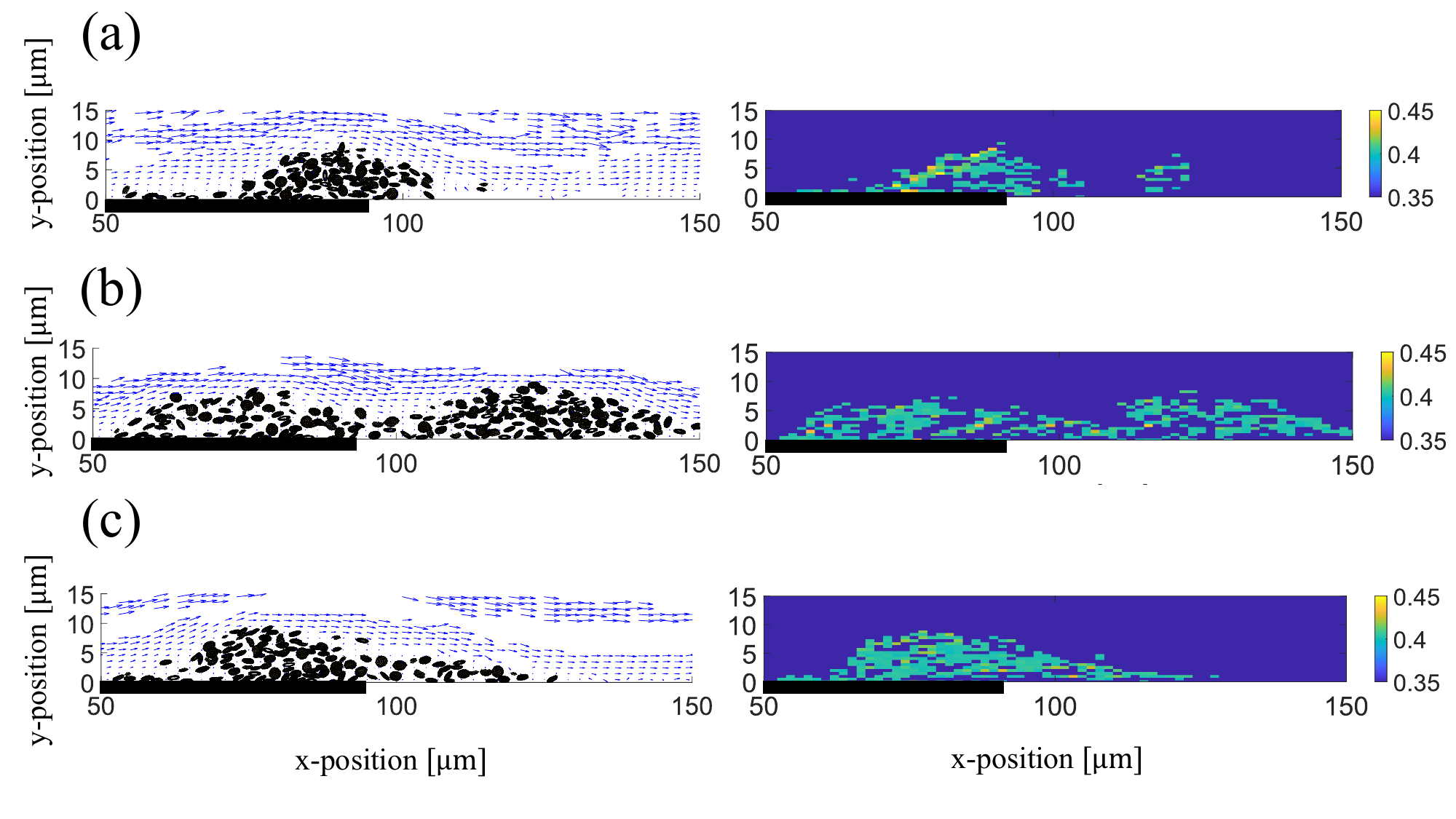}
  \captionsetup{format=plain}
  \caption{Structure of the clot during embolization events, illustrated by the conformation in a two-dimensional ($xy$-plane) slice, obtained from the $z=15\mu m$
  at various times.  The black bar depicts the location of the hemostatic surface in all figures. (a) \textit{Formation} phase of platelet-vWF plug development. 
  The flowing platelets and vWFs are captured by the hemostatic surface and form a growing plug at the top of the wound.
  (b) \textit{Extension} phase with displacement of platelets downstream from the vascular injury due to hydrodynamic forces. 
  The two distinct parts are still connected by vWF-platelet chains. (c) \textit{Fracture} phase of platelet-vWF plug. The second 
  part at the downstream is now detached from the clot. At later times, this configuration evolves to a similar configuration as in (a). 
  (Left column) Conformation of platelets attached to the clot, and fluid velocity field (blue arrows) during cyclic \textit{formation-extension-fracture} 
  process. (Right column) Heat map of the extension of the vWF-platelet bonds. The largest extension of vWF-platelet bonds occurs in (a) 
  at the leading edge of the clot, and in (b) between the vWF-platelet bonds connecting the two clot parts. This process is 
  illustrated in Movies~S1, S2, and S3 for the low-, mid-, and high-flow-rate simulations, respectively.}
  \label{fig:embolization}
\end{figure*}

The mechanism of embolization is illustrated in Fig.~\ref{fig:embolization} for the low-flow-rate clot (see also Movie S1), 
using a slice parallel to 
the $xy$-plane that cuts the simulation domain exactly in half along the $z$-axis. The mechanism is identical 
for the clots for mid- 
and high-flow rates (see Movies S2 and S3), and is therefore depicted only for the low-flow-rate case. Figure \ref{fig:embolization} displays the platelets 
within the clot (left column) as well as local extensions of the platelet–vWF bonds. In the initial growth phase [Fig.~\ref{fig:embolization}(a)],
the clot is located primarily at the top of the wound and the highest bond force between vWF-platelet pairs is localized near the leading edge 
of the clot. This early phase is referred to as \textit{formation} phase. As the clot grows, the drag force due to fluid flow increases, such that 
a part of the clot is pushed toward the downstream direction. This stage (referred to as \textit{extension} phase) is shown in Fig.~\ref{fig:embolization}(b), 
where some portion of the clot is still at the top of the wound, while the other part is extended downstream near the lower wall of the channel. 
The part located above the wound is firmly anchored to the hemostatic surface, whereas the free part beyond the wound is slowly extended 
further until a neck is formed between the anchored and free parts. At this stage, the largest stretch of the platelet-vWF bonds is located 
in the neck [see the right column of Fig.~\ref{fig:embolization}(b)]. Eventually, the neck thins and the free part detaches from the clot, 
resulting in embolization (referred to as \textit{fracture} phase), as displayed in Fig.~\ref{fig:embolization}(c).  
This cycle of clot growth, extension, and embolization continues for all studied flow rates. As a result, the \textit{formation-extension-fracture} process
is the embolization pathway in primary hemostasis. In all our simulations, detachment of clot fragments from the wound zone is negligible, 
and the embolization primarily comes from the freely extended part of the clot. 

\begin{figure*}[t]
  \centering
  \includegraphics[width=1\textwidth]{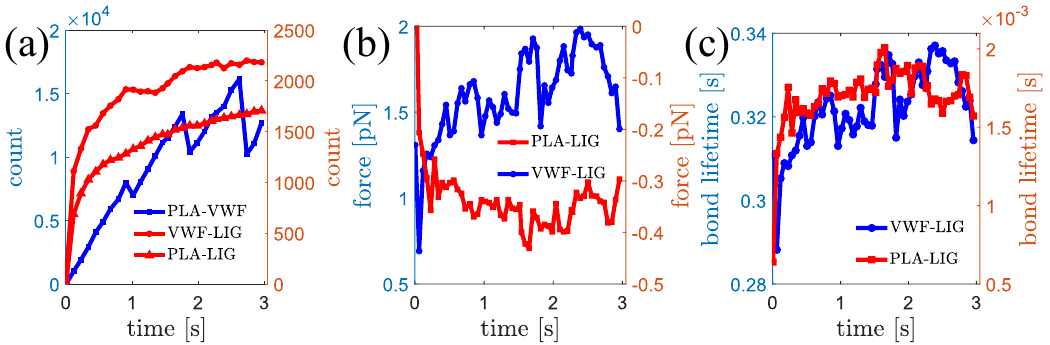}
  \captionsetup{format=plain}
  \caption{Bond properties for the low-flow rate clot. (a) Number of bonds for different interaction pairs as a function of time. (b) Ensemble 
  average of bond lifetimes corresponding to different pairs. (c) Ensemble average of the forces between vWF-ligand and platelet-ligand pairs 
  as a function of time.}
  \label{fig:bondAnalysis}
\end{figure*}

\subsection{Clot formation -- internal bond structure and forces}

Within the clot, there are different adhesive bonds that sustain structural forces and attain a certain bond lifetime, shown in 
Fig.~\ref{fig:bondAnalysis} for the low-flow rate clot. Note that the results for mid- and high-flow-rate simulations are not presented, 
because the behavior is similar to the low-flow-rate case. Since the wound has a finite size, there is a limited number of available 
ligands (i.e., total of $19476$ ligands) to form bonds with the hemostatic surface. As the simulation progresses, these ligands are quickly 
occupied by bound vWFs and platelets. The ratio of number of vWF-ligand and platelet-ligand bonds is nearly constant throughout the simulation 
and equals $0.75$. Only about $18\%$ of the ligands are occupied within the steady-state time $\tau_{\rm ss} \approx 1.18 \rm s$, beyond which the number of bonded ligands increases linearly in time, as shown in Fig.~\ref{fig:bondAnalysis}(a). 

Inside the clot, the number of bonds between vWFs and platelets is proportional to the clot size (i.e., total number of components in the clot).
For a clot after $\tau_{\rm ss}$, $50\%$ of the monomers of the attached vWFs are activated, with a majority of them ($\approx 90\%$) 
occupied with either ligand-vWF or platelet-vWF bonds. As noted earlier, when embolization occurs, a clot fragment detaches from the trailing 
edge, and not from the anchored zone. Therefore, the number of bonds with ligands is not affected, whereas the number of vWF-platelet bonds 
is subject to sudden jumps, as presented in Fig.~\ref{fig:bondAnalysis}(a).

Once the clot is formed, it is anchored to the wound by ligand-vWF and ligand-platelet bonds. Each bond is loaded depending on the hydrodynamic 
drag force applied to the blood clot. Figure~\ref{fig:bondAnalysis}(c) shows the force carried by each bond averaged over all bonds at a particular 
time step. Interestingly, the platelet-ligand bonds are compressed on average, whereas vWF-ligand bonds are stretched. Note that both tensile 
and compressive forces appear throughout the simulation for both bond types and different flow rates. The force carried by platelets on average 
almost vanishes. Thus, the resistance of the clot against hydrodynamic forces is carried predominantly by the bonds of ligand-vWF pairs.

The force carried by the bond is directly related to the corresponding bond lifetime. Figure~\ref{fig:bondAnalysis}(b) shows the respective bond 
lifetimes for different bonds with the hemostatic surface. The bond lifetime for platelet-ligand interactions is short (of the order of $10^{-3}s$),
whereas the lifetime of vWF-ligand bonds is two orders of magnitude larger (of the order of $10^{-1}s$). This is a further confirmation that 
platelet-ligand interactions during primary hemostasis for high-flow-rate environments are almost negligible for clot attachment to the wounded 
area. 

\begin{figure*}[t]
  \centering
  \includegraphics[width=1\textwidth]{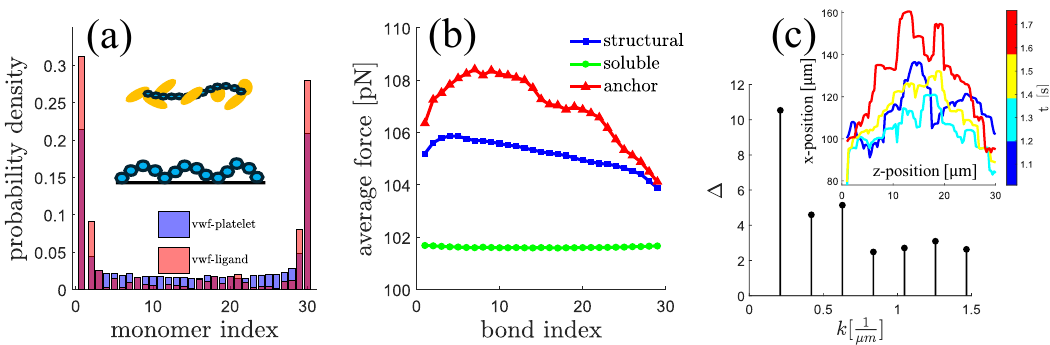} 
  \captionsetup{format=plain}
 \caption{Bond characteristics along vWF chains and the clot shape. (a) Probability of bond formation between the active monomers of vWF chains 
 along their contour length with platelets or ligands. (b) Ensemble-averaged forces derived from the bond potentials [Eqs.~\eqref{eq:FENE} and 
 \eqref{eq:lj}] between the neighboring monomers of different types of vWF chains. Structural vWFs are part of the clot without forming adhesive 
 bonds with ligands at the hemostatic surface. Anchored vWFs have at least one monomer attached to the wound surface. Soluble vWFs do not participate 
 in any adhesive interactions. (c) Amplitudes $\Delta_{i} = \sqrt{A_{i}^{2} + B_{i}^{2}}$ from the Fourier spectral analysis
 [see Eq.~(\ref{eq:fourier})] of clot edge profiles for the first six wavenumbers $k_i=2\pi i/L_z$. The inset displays the 
 $xz$ projection of the ensemble-averaged clot edge.}
 \label{fig:geometricProperties}
\end{figure*}


Figure~\ref{fig:geometricProperties}(a) shows the probability distribution of bond formation between vWF monomers and their binding partners 
(i.e., either platelets or ligands) along the vWF contour for the low-flow-rate case. A single vWF chain attaches to the hemostatic surface 
mainly at one of its ends. Within the clot, vWF chains assume a snake-like shape, where intermediate parts of the chain form occasional 
attachments with the hemostatic surface. vWF-platelet interactions are similar to vWF-ligand interactions, but intermediate monomers 
have a higher probability to form adhesive bonds. The two adhesion conformations are illustrated schematically in the inset of
Fig.~\ref{fig:geometricProperties}(a). 

The average-force distributions along vWF chains are shown in Fig.~\ref{fig:geometricProperties}(b). Soluble chains, which do not participate 
 in any adhesive interactions, exhibit an almost uniform tension profile. In contrast, when a vWF chain is adhered to the hemostatic surface 
 (i.e., anchored chain) or forms a bond with a platelet (i.e., structural vWF), the average tension first increases, and then decreases along 
 the contour. Furthermore, the tension is largest at the front (adhered) end of a vWF chain, resembling the tension distribution of a polymer 
 attached to a surface and subjected to a uniform external load along its monomers (so that all forces sum up at the front end, but the drag 
 force of only a few monomers remains at the trailing end). Within the clot, this ideal tension distribution is affected by adhesive bond 
 formation in the middle and end of the chain, which implies a reduced tension near the trailing end. Furthermore, the clot is a three-dimensional 
 elastic body, anchored to the wall. This means that the external flow is screened, and flow forces are weak inside the clot, while the main drag 
 forces occur at the free surface of the clot. Thus, the highest forces can be expected for vWF–ligand interactions at the wall, while 
 the smallest forces are near the top surface of the clot.

\subsection{Instantaneous clot geometry and shape}

We also measure the aspect ratio of the clot by comparing its lengths $\ell_x$ and $\ell_y$ along the corresponding axes. The length of the 
clot in flow direction, $\ell_{x}$, can become as large as twice the length of the hemostatic surface, or about $100 \mu m$. The dimension 
$\ell_{y}$ is the peak height, which is displayed in Fig.~\ref{fig:surfaceCoverage} for different flow-rate cases. Since the clot is subject 
to embolization, $\ell_{x}$ and $\ell_{y}$ experience sudden jumps. Despite these events, the aspect ratio of the clot remains nearly constant 
between two jumps, which implies that clot rupture reduces the size significantly, but the remaining clot quickly recovers its previous size 
and shape. The ratio remains $\ell_y/\ell_x = 0.12 \pm0.01$, once the time $\tau_{\rm ss}$ is reached.

To characterize the shape of the trailing edge, Fig.~\ref{fig:geometricProperties}(c) shows the Fourier spectrum of the projection of edge 
contour $x(z)$ of the clot onto the $xz$ plane. The inset in the same figure displays the corresponding averaged boundary curve after 
$\tau_{\rm ss}$ for the low-flow-rate simulation between two embolization events. The Fourier representation of the contour is given by 
\begin{equation}
\label{eq:fourier}
x(z) = \sum_{i=1}^{N_{\rm mode}} \left[
A_{i}\cos\left(k_i z\right) 
+ B_{i}\sin\left(k_i z\right)
\right], 
\end{equation}
where $k_i=2\pi i/L_z$ is the wavenumber of the $i$th Fourier mode, $A_{i}$ and $B_{i}$ are the corresponding coefficients. The rapid drop in amplitude with increasing wavenumber
$k_i$ implies that large wavenumbers (or short wavelengths) are not significant, so that the contour is smooth and dominated by one or two mesoscopic length 
scales --- consistent with the “tongue-like” elongation shown in the inset of Fig.~\ref{fig:geometricProperties}(c). This specific "tongue-like" 
shape appears due to material flow. As clot grows, the material on the top of the clot is slowly dragged by hydrodynamic forces toward 
the trailing edge (see Movies S1, S2, and S3).

\section{Discussion and conclusions}

We have investigated the formation and dynamics of a vWF-platelet clot at a vascular injury in blood flow for physiologically relevant 
flow rates. For all considered flow conditions, clot formation begins with the attachment of vWF chains and platelets at various locations 
on the wound surface, creating small discrete islands. These islands gradually merge into a single connected blood clot that covers 
a large part of the wound and in all cases, extends beyond the hemostatic surface. Even though the structural properties (e.g., total number 
of components, bond counts, surface coverage) of the clots are similar, the dynamics is a strong function of the flow rate. As the clot grows 
with time and partially occludes the channel, the resulting increase in hydrodynamic forces pushes portions of the clot beyond the wound site, 
leading to the detachment of fragments of the clot. This process continues cyclically as long as the supply and transport of hemostatic 
material (e.g., vWFs and platelets) are maintained.

The behavior of clot formation for the range of applied flow rates is as follows. The steady-state time $\tau_{ss}$ is inversely proportional 
to wall-shear rate $\dot{\gamma}_{\rm wall}$ of the flow, and scales as $\tau_{ss} = 1.3 \times 10^{3}/\dot{\gamma}_{\rm wall}$. The hemostatic 
surface coverage at a time $\tau_{ss}$ after flow initiation reaches approximately same values for all flow rates, and fluctuates within 
the range of $58-76\%$ of the total wound area. This area measure is not affected by the embolizations from the clot, i.e., as soon as 
a steady-state clot is reached, the primary hemostasis is essentially completed, and the system is ready to proceed with the secondary 
hemostasis without interruption. The material deposition rate (i.e., total number of vWFs and platelets per second) is proportional 
to applied flow rate with a scaling of $\dot{N} \approx 1.79 \dot{\gamma}_{\rm wall}$. The composition of the clot is homogeneous and 
keeps about the same vWF-to-platelet number ratio of approximately $0.55$, as in the periodic inflow domain.   

It is important to note that, even though the hemostatic surface coverage is free of disruptions, under physiologically relevant flow conditions 
with whole blood, embolization occurs through a three-step process: clot formation -- flow-induced extension toward the downstream side -- rupture and 
embolization. Due to limited time covered by our simulations, a quantitative estimate of the average embolization time was not possible. However, 
the observed mechanism of primary hemostasis suggests that the transition to secondary hemostasis is initiated in the vicinity of the wound. 
Fibrin bridges could form gradually from platelets at the vascular surface to the platelets at the upper portion of the clot, thus stabilizing 
the clot from bottom to top, until the clot solidifies and embolization is prevented.

Previous stability studies \cite{Nechipurenko_BJ_2024,Ataullakhanov_JTB_2013} of blood clots formed under physiologically relevant flow have typically assumed a predefined clot shape based on 
experimental observations and distinguished between weak and strong platelet–platelet bonds. In these studies, embolization most often occurs 
at the free surface of the clot, where the local shear rate is highest. In our simulations 
without platelet activation, the upper part of the clot is continuously displaced away from the wound location, while maintaining a finite 
amount of hemostatic material at the wound site. We therefore predict that secondary hemostasis should be initiated within this retained material, 
which is not subjected to continuous scavenging. Once the material adjacent to the wound solidifies, subsequently deposited material can 
accumulate on this solidified scaffold and undergo secondary hemostasis. In this manner, the clot can grow and solidify progressively in 
a layer-by-layer fashion. 

A crucial result of our simulations is the finite size of the clot. We model clot formation in the presence of RBCs, platelets, and vWF molecules, 
but in the absence of soluble fibrinogen, using a hemostatic surface that allows vWFs and platelets to attach. Also, our model does not include 
any biochemical reactions that produce additional fibrinogen at the injury site. Under these conditions, we observe no channel occlusion, 
unlike recent experiments \cite{Ku_JTH_2016}. A key difference is the presence of RBCs in our simulations, which exert additional drag 
on the clot and contribute to its scavenging. Under whole-blood flow conditions with blocked fibrin polymerization, clots embolize and 
the probability of occlusion is low \cite{Diamond_AVTB_2012}, consistent with our findings.

To characterize the micro-environment of the clot, we also measured the average minimum distance between platelets inside the clot. 
Recent simulations and experiments \cite{Brass_Core1_2014,Brass_Core2_2014} indicate that after the formation of platelet plug, platelets 
form fibrin bridges and these bridges contract to decrease the porosity of the environment. In this micro-environment, the material transport 
favors diffusion over advection, so that the chemicals released to initiate/propagate secondary hemostasis are not scavenged by the fluid 
flow. We have computed local inter-platelet distances from our simulations in the absence of platelet activation and without the formation 
of fibrin bridges, when the peak height of the clot reaches about $12\mu m$. Next to the wall, all averaged distances are nearly 
the same for all studied flow rates and correspond to a minimum inter-platelet distance of approximately $0.18\mu m$. This minimum 
value then linearly rises up to $0.21 \mu m$ until the distance to the wall becomes $10 \mu m$. In the outermost $2\mu m$-thick layer, 
this distance increases exponentially to $0.35\mu m$. These results support the concept of distinct core and shell regions, where 
the core’s shielding effect favors diffusion-dependent material transport and thereby, facilitates secondary hemostasis. 
No net dependence of the minimum inter-platelet distance on the flow rate is observed. The lowest distance is about $0.17\mu m$ 
next to the wall for mid-flow rate, whereas for high-flow-rate simulation this distance is $0.19 \mu m$ next to the wall. 

In conclusion, this work presents a direct numerical framework for investigating the long-time dynamics of primary hemostasis under 
physiologically relevant flow conditions. By explicitly resolving RBCs, platelets, and shear-activated vWF, the model captures 
the mechanically driven processes that govern platelet adhesion, aggregation, and the transport of hemostatic material during 
the early stages of clot formation. The simulations highlight the crucial role of hydrodynamic interactions and shear-dependent 
mechanisms in regulating clot growth, stability, and embolization, providing details that are difficult to access experimentally. 
Building on this foundation, it would be particularly interesting to extend the present model by incorporating fibrin generation 
and polymerization, thereby enabling a unified {\it in silico} description of the transition from primary to secondary hemostasis 
and the subsequent mechanical stabilization of the clot.

\section*{Author Contributions}

G.G. and D.A.F. conceived the research project. A.T. performed the simulations and analysed the obtained data.   
All authors participated in the discussions and writing of the manuscript.

\section*{Conflicts of interest}

The authors have declared that no competing interests exist.

\section*{Acknowledgements}

The authors gratefully acknowledge computing time on the supercomputer JURECA \cite{jureca} at Forschungszentrum 
J{\"u}lich under grant no. actsys.

\section*{Data availability}

The data that support the findings of this article are openly available \cite{data_rep}.

\appendix  

\section{Modeling fluid flow and boundary conditions}
\label{sec:app_model}

In the smoothed dissipative particle dynamics (SDPD) approach \cite{Espanol_SDPD_2003, Mueller_SDPD_2015}, fluid 
is represented by a collection of interacting particles, each corresponding to a finite fluid volume. Interactions between particles 
include conservative, dissipative, and stochastic forces. The conservative component controls fluid compressibility by enforcing 
an equation of state of the form $p = p_0 (\rho / \rho_0)^\gamma - b$, where $\rho$ is the local particle density, $\rho_0$ is a 
reference density, and $p_0$, $\gamma$, and $b$ are parameters that determine the fluid’s compressibility and set the speed of sound 
via $c^2 = p_0 \gamma / \rho_0$. In all simulations, the SDPD fluid parameters are $p_0 = b = 33750 k_BT/r_c^3$, $\gamma = 7$, 
$\rho_0 = mn$ with particle mass $m$ ($m=1$ in simulations) and number density $n$. Here, $r_c$ is the cutoff radius that limits 
the range of all fluid forces. Fluid viscosity is controlled by the dissipative interactions, while the balance between dissipative 
and random forces maintains the system at a fixed temperature, effectively acting as a thermostat. 
In this study, we adopt an SDPD formulation that conserves angular momentum, in addition to mass and linear momentum, as 
described in Ref.~\cite{Mueller_SDPD_2015}.

To enforce no-slip boundary conditions at the channel walls, we embed a layer of immobile (frozen) particles, with a thickness equal 
to the interaction cutoff radius $r_c$. These wall particles are structurally consistent with the bulk SDPD fluid, exhibiting 
the same pair distribution function. Their role is to provide the necessary interaction forces to nearby fluid particles and 
thereby mimic a solid boundary. To prevent fluid particles from penetrating the wall, a bounce-back reflection is applied at 
the fluid–solid interface. However, because standard dissipative interactions alone may not sufficiently maintain a strict 
no-slip condition, an adaptive shear force is additionally applied to fluid particles located within a near-wall region of 
thickness $r_c$ \cite{Fedosov_TDI_2009, Lei_TDO_2011}.

Interactions between the fluid and suspended biological components --- such as blood cells and vWF molecules --- are captured 
by a dissipative coupling, which ensures momentum exchange and mimics viscous drag. This approach is analogous to the coupling 
mechanism used in the immersed boundary method \cite{Fedosov_RBC_2010, Ahlrichs_SSC_1999}, allowing for efficient integration 
of deformable structures into the SDPD framework.

Periodic boundary conditions often used in particle-based simulations can lead to material depletion when there is a sink present 
within the simulation domain. In our case, the vascular injury acts as an active sink, continuously capturing hemostatic material 
from the free flow, as the simulation progresses. To prevent such depletion, we implement non-periodic boundary conditions at the inflow 
and outflow ends of the channel, as illustrated in Fig.~\ref{fig:model}(c). In this setup, the simulation domain is divided into 
two subdomains (i.e., inflow and observation domains) separated by a virtual interface. The inflow domain implements periodicity between its 
ends, and at the same time supplies the flow of blood to the observation domain (see Fig.~\ref{fig:model}(c)) \cite{Lykov_IOC_2015}, 
where the hemostatic reaction occurs. The observation domain is bounded at the downstream by an outflow boundary, where particles which 
leave the simulation domain are removed \cite{Lei_TDO_2011,Lykov_IOC_2015}. To maintain continuity of mass and force balance at the 
outflow boundary, a SDPD pressure-force is applied to the fluid next to outflow within a cut-off radius \cite{Lei_TDO_2011}. 

\section{Model of shear-activated vWF}
\label{sec:app_vwf}

In the bead-spring model of vWF, monomers are linked by finite-extensible nonlinear-elastic (FENE) 
springs, allowing for realistic stretching behavior under flow, with interaction potential
\begin{equation}
U_{\rm FENE}(r) = -\frac{k_{\rm s}}{2} r_{\rm max}^2 \ln \left( 1- \left( \frac{r}{r_{\rm max}} \right)^2 \right). \label{eq:FENE}
\end{equation}
The spring connecting two adjacent monomers has a stiffness $k_{\rm s}$, and its extension is limited to a maximum of 
$r_{\rm max} = 2\sigma$, where $\sigma = 0.077 D_r$ is the diameter of a monomer. The variable $r$ denotes the instantaneous 
distance between neighboring monomers. In all simulations, the spring constant is set to $k_{\rm s} = 25000k_{\rm B}T / \sigma^2$, 
ensuring that the springs behave as effectively inextensible connectors. Excluded volume effects and weak self-attraction between 
monomers are taken into account using a 12-6 Lennard-Jones (LJ) potential 
\begin{equation}
U_{\rm LJ}(r) = 4\epsilon \left[ \left(\frac{\sigma}{r} \right)^{12} -  \left(\frac{\sigma}{r} \right)^{6} \right],
\label{eq:lj}
\end{equation}
where $\epsilon=16k_{\rm B}T$ controls the attraction strength. The LJ interaction for monomers within the same
vWF is cut off beyond a distance $r_{LJ} = 2.5 \sigma$. vWF chains are coupled to the SDPD fluid using a frictional
force \cite{Roemer_DBP_2015, Huisman_ADAM_2017}. 

This vWF model exhibits shear-dependent conformational properties, maintaining a compact globular shape at low shear rates and 
undergoing stretching under sufficiently high shear, consistent with experimental observations 
\cite{Katz_FIU_2006, Katz_DIP_2008, Huisman_ADAM_2017, Hoore_FIA_2018, Schneider_SIU_2007, Reininger_VWF_2008, Fu_VWF_2017}. 
Upon stretching, vWF exposes binding sites, enabling it to interact with specific receptors on platelet surfaces. The onset of adhesiveness 
in the model is governed by the local conformation of the chain, and is determined by two geometrical criteria that describe 
monomer-level deformation \cite{Huisman_ADAM_2017}. The first criterion is based on the bond angle formed by three consecutive 
monomers $i-1$, $i$, and $i+1$, defined as $\theta_{i-1,i,i+1}$. A monomer is considered stretched --- and therefore eligible 
for activation --- if, $\theta_{i-1,i,i+1} \ge \theta_{\rm thres}$, $2 < i < N_m - 1$. The threshold angle $\theta_{\rm thres}$ 
quantifies the degree of local elongation required for activation. The angle criterion is automatically satisfied for the first 
and last monomers in the chain. The second criterion ensures that the monomer is not embedded in a globular region of the chain. 
It checks for the absence of non-adjacent neighbors within a threshold radius, requiring $N_{neigh}(r_{ij} \ge R_{\rm thres},\, 
j \neq i,i\pm 1) = 0$. This condition prevents activation if the monomer is spatially close to distant parts of the chain. 
In all simulations, we set $\theta_{\rm thres} = 150^\circ$ and $R_{\rm thres} = 1.2\,\sigma$, following prior studies 
\cite{Hoore_FIA_2018}, where the sensitivity of vWF activation to these parameters has been analyzed. If an activated monomer 
forms a bond with a platelet or hemostatic surface, then the above criteria become ineffective --- making the monomer to remain 
active till the bond is broken. In summary, a monomer becomes activated only if both criteria are satisfied simultaneously and 
stays active whenever it remains bound to a platelet or hemostatic surface, even if one or both of the criteria are violated. 
If an active monomer becomes unbound, it is deactivated as soon as either condition is violated.

\section{Description of movies}

\begin{itemize}
    \item {\bf Movie S1}. Evolution of a hemostatic plug at the wound for low flow rate 
    (wall shear rate $\dot\gamma_w=1152 s^{-1}$). Platelets are shown in yellow, 
    while inactive and active VWF monomers are drawn in blue and green, respectively. Accumulating hemostatic material initially forms 
    a single island at the wound (i.e., formation phase). Then, hydrodynamic drag forces push the upper layer downstream, leading to a 
    significant extension of the clot (i.e., extension phase). Finally, a portion of the clot detaches from the main plug, resulting 
    in embolization (i.e., fracture phase). The formation-extension-fracture process repeats again and again.  
    
    \item {\bf Movie S2}. The formation-extension-fracture process for medium flow rate (wall shear rate $\dot\gamma_w=1728 s^{-1}$). RBCs are shown in red, the other components 
    are drawn by the same colors as in Movie S1.  
    
    \item {\bf Movie S3}. Dynamics of a hemostatic plug at the wound for high flow rate (wall shear rate $\dot\gamma_w=2304 s^{-1}$). A qualitatively similar 
    formation-extension-fracture process is observed in comparison to the cases in Movies S1 and S2. The color code is the same 
    as in Movie S1.
\end{itemize}

\bibliography{main}

\end{document}